\documentclass[pra,aps,twocolumn,eqsecnum,showpacs]{revtex4}

\usepackage{dcolumn}
\usepackage{amsmath}
\usepackage{graphicx}
\usepackage{epstopdf}

\begin{document}

\title{Quantum memory for light via stimulated off-resonant Raman process:\\beyond the three-level $\Lambda$-scheme approximation}

\author{A.S. Sheremet, L.V. Gerasimov, I.M. Sokolov, D.V. Kupriyanov}%
\affiliation{Department of Theoretical Physics, State
Polytechnic University, 195251, St.-Petersburg, Russia}%
\email{kupr@dk11578.spb.edu}%

\author{O.S. Mishina, E. Giacobino, J. Laurat}%
\affiliation{Laboratoire Kastler Brossel,
Universit\'{e} Pierre et Marie Curie, Ecole Normale Sup\'{e}rieure,
CNRS, Case 74, 4 place Jussieu, 75252 Paris Cedex 05, France}%

\date{\today }

\begin{abstract}
We consider a quantum memory scheme based on the conversion of a
signal pulse into a long-lived spin coherence via stimulated
off-resonant Raman process. For a storing medium consisting of
alkali atoms, we have calculated the Autler-Townes resonance
structure created by a strong control field. By taking into
account the upper hyperfine states of the $D_1$ optical
transition,  we show important deviations from the predictions of
the usual three-level $\Lambda$-scheme approximation and we
demonstrate an enhancement of the process for particular detunings
of the control. We estimate the memory efficiency one can obtain
using this configuration.
\end{abstract}


\maketitle%

\section{Introduction}
Long-lived and highly efficient quantum memories for light are a
crucial tool for quantum information processing, quantum
computing, long-distance secure communication and scalable
networks \cite{Kimble,HammSorPol}. Various physical systems are
intensively investigated and many significant advances have been
recently reported \cite{QMemories,Lvovsky}. In particular, large
ensembles of identical atoms, as gases at room temperature
\cite{Eisaman,NovLuk,COBPG,AppLvov,HetetLam} or ultracold samples
\cite{ChanKuz,CDLK,HonKoz}, have been successfully used for
demonstrating the storage and retrieval of quantum states in
different regimes.

In the theoretical investigations of such quantum memories, the
description of the light-atoms interface is usually based on a
three-level $\Lambda$ configuration, with two ground states and
one excited state \cite{Gorshkov,NunnJak,MLSSK}. However, the
hyperfine interaction in the upper states of alkali atoms is not
strong enough for the system to be perfectly described by this
approximation. The multilevel structure of alkali atoms has thus
to be taken into account for a correct description of the quantum
memory effect  \cite{MKMP,MSLK}. In this paper, we show for the
$D_1$ optical transition how an additional level significantly
modifies the stimulated Raman process and we finally evaluate the
efficiency of pulse storage and retrieval in this configuration.

The paper is organized as follows. Section \ref{SectionII} first
introduces the energy diagram and the assumptions on which our
work is based. We then present the theoretical model and provide a
general expression of the atomic sample susceptibility. In Section
\ref{SectionIII}, we study the susceptibility behavior for
different frequency detunings of the control and probe modes and
we explain how the multilevel structure significantly modifies the
quantum light transport driven by coherent scattering mechanism in
the conditions of electromagnetically-induced transparency or
stimulated Raman process. In section \ref{SectionIV}, we present
calculations for the transport of a signal pulse in an optically
dense sample. The signal pulse is tuned in the spectral region
where the Autler-Townes resonance would be maximal in its
amplitude and the stimulated Raman scattering would be most
effective. The achievable efficiency of a quantum memory protocol
performed in this configuration is finally discussed, in both
cases of forward and backward retrieval.

\section{Autler-Townes effect in the $D_1$-line of alkali
atom}\label{SectionII} In this section, we present the energy configuration
under study and we detail the theoretical model used to fully describe this
configuration. General expression of the susceptibility of the medium is
finally given.

\subsection{Basic assumptions}

\begin{figure}
\includegraphics[width=0.5\textwidth]{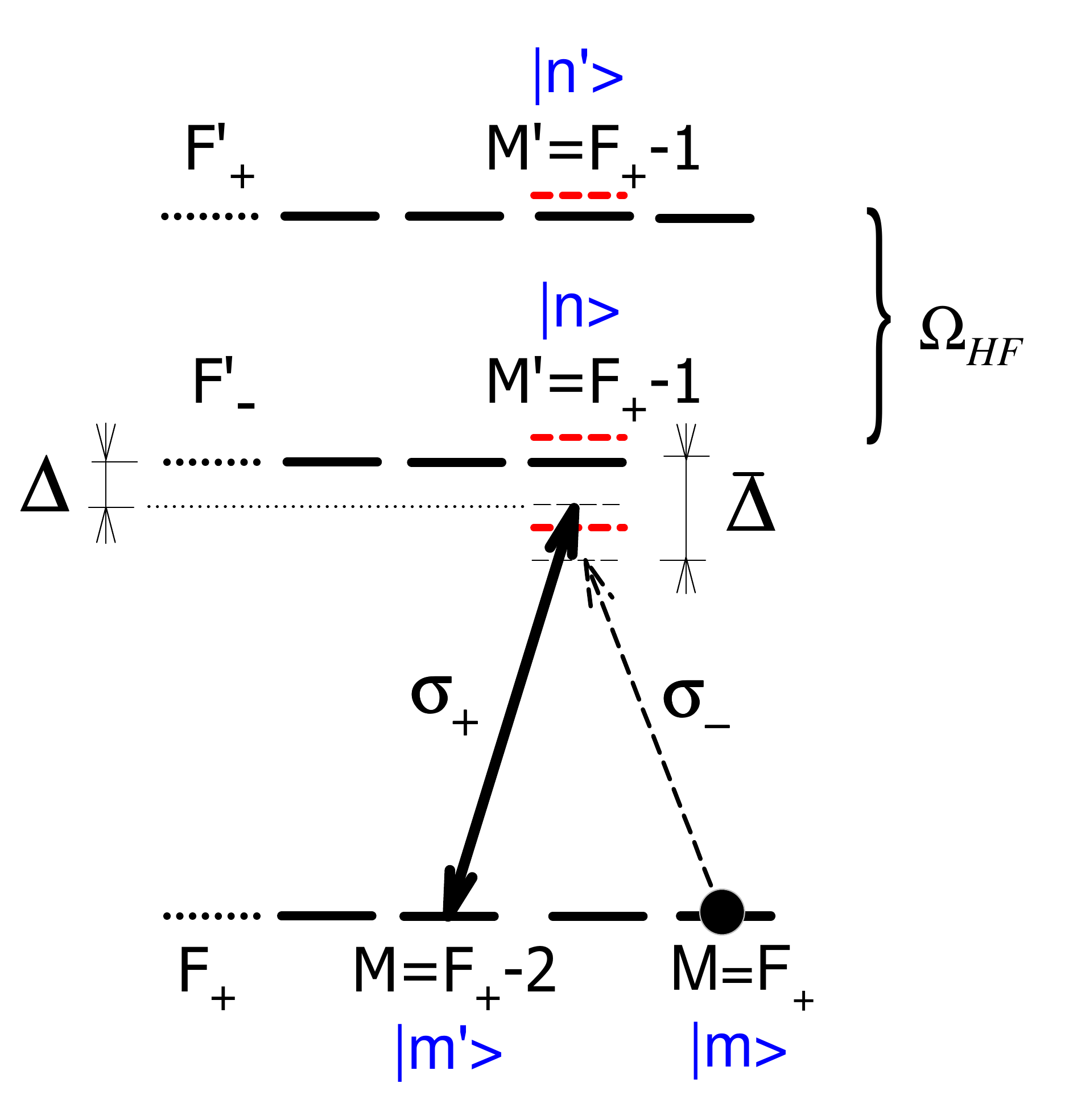}%
\caption{(Color online) Energy levels and excitation channels
considered for the $D_1$ line of alkali atom. The atoms populate
the upper hyperfine sublevel of their ground state. The system is
dressed by a strong control field with a right-handed circular
polarization ($\sigma_+$) and a frequency detuning $\Delta$. The
system is probed by a weak mode with left-handed circular
polarization ($\sigma_{-}$) and frequency detuning $\bar{\Delta}$.
The three Autler-Townes resonances (AT-triplet) are shown by the
red dashed lines.}
\label{fig1}%
\end{figure}

In this paper we consider the case of alkali atoms and focus in
particular on the $D_1$ optical transition. Figure \ref{fig1}
gives the energy diagram and figure \ref{fig2} shows the
excitation geometry. The atoms populate the upper hyperfine
sublevel of their ground state with maximal spin projection
$\left\{F_{+}=I+1/2, M=F_{+}\right\}$ (where $I$ is atomic nuclear
spin) and we denote this state as $|m\rangle$.

The control field has a right-handed circular polarization
($\sigma_+$). In this configuration, there is no interaction of
the control field with the populated sublevel and only the
presence of the probe mode in the orthogonal left-handed circular
polarization ($\sigma_-$) opens the excitation channel. The probe
and control modes coherently couple the populated state
$|m\rangle$ with the Zeeman sublevel $\left\{F_{+},
M=F_{+}-2\right\}$ in the ground state, which we denote as
$|m'\rangle$, and with two Zeeman sublevels in the excited state
$\left\{F'_{-}=I-1/2, M'=F_{+}-1\right\}$ and
$\left\{F'_{+}=I+1/2, M'=F_{+}-1\right\}$, which we denote
respectively $|n\rangle$ and $|n'\rangle$.

Such a configuration leads to the Autler-Townes (AT) effect
\cite{AutlerTownes,LethChebt}. As shown in figure \ref{fig1}, the
energy levels of the excited atomic states $|n\rangle$ and
$|n'\rangle$ are modified by the presence of the control field. If
the control field is tuned precisely in resonance with the
non-disturbed transitions, either $|n\rangle$ or $|n'\rangle$, the
interaction with the control mode splits the originally degenerate
system of the atomic and field oscillators into two quasi-energy
sublevels. However, the interaction of the control with the other
upper hyperfine atomic state is not negligible and the presence of
the control mode affects on the locations of both the hyperfine
resonances.  For an arbitrary detuning of the control, we refer to
the joint atom-field quasi-energies as the AT-triplet (instead of
the usual AT doublet) resonance structure. The location of these
three resonances for a particular detuning of the control mode is
shown by the dashed red lines in figure \ref{fig1}.

Let us note finally that the $\Lambda$-scheme approximation is the
asymptotic limit of our approach: in this limit, the hyperfine
interaction is considered to be much stronger than either the
natural decay rate $\gamma$, the Rabi frequency of the control
mode or the detuning $\Delta$. We will show the limit of this
widely-used approximation via the full calculation of the AT
structure for the $D_1$-line of ${}^{133}$Cs atom, which has the
largest hyperfine splitting among
 alkali atoms $\Omega_{HF}=1168$ MHz $=256\,\gamma$.

\subsection{Calculation approach}

\begin{figure}
\includegraphics[width=0.5\textwidth]{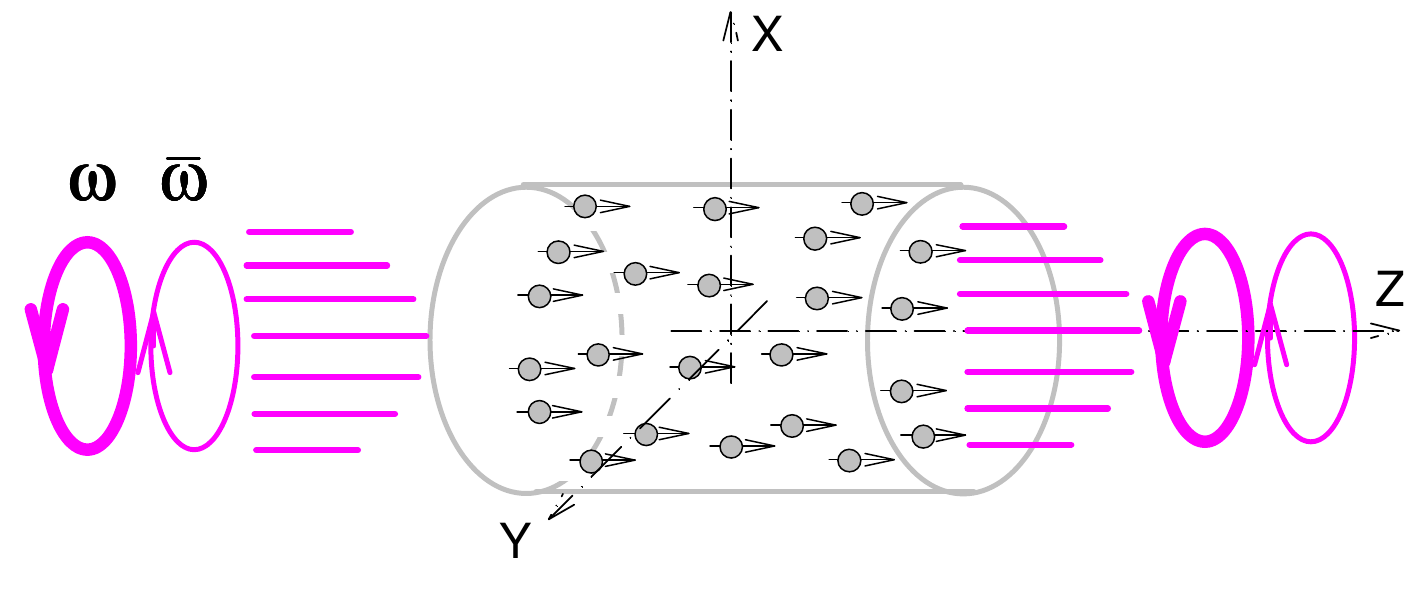}%
\caption{(Color online) Schematic diagram showing the excitation geometry for
the transitions specified in figure \ref{fig1}.}
\label{fig2}%
\end{figure}

We now turn to our theoretical model. The AT resonance structure
can be observed by passing through the sample a probe pulse that
will excite the atoms from the populated state $|m\rangle$. The
propagation of the probe mode through the medium, in the geometry
shown in figure \ref{fig2}, is described by the standard
macroscopic Maxwell equation, where the dielectric susceptibility
of the medium, being linear in response to the probe mode,
accumulates all the orders of nonlinearities for the control mode.
Due to the linearity for the probe field, we can consider an
arbitrary set of probe field frequencies forming the signal pulse
and we follow the temporal and spatial dynamics of this pulse in a
finite sample. The dynamics of the signal pulse and control mode
come from the characteristics of the stimulated Raman process, as
shown in \cite{LethChebt,Wang}. The absorption and dispersion
parts of the AT-resonances in the sample susceptibility result
from the combined action of this process on the signal pulse
transport.

The positive frequency component ${\cal E}^{(+)}(\mathbf{r},t)$ of the signal
pulse can be described by a wavepacket propagating along the direction $z$,
with a carrier frequency $\bar{\omega}$ and wavenumber
$\bar{k}=\bar{\omega}/c$, such as:
\begin{equation}
{\cal E}^{(+)}(\mathbf{r},t)=%
\epsilon(z;t)\mathrm{e}^{-i\bar{\omega}t+i\bar{k}z}.%
\label{2.1}%
\end{equation}
For the sake of simplicity the transverse profile of the
wavepacket is not taken into account, and we thus ignore any
diffraction and transverse inhomogeneity of the atomic sample
during the propagation process. However let us point out that the
analysis could be generalized for an inhomogeneous system as well.

We then make use of the Fourier expansion of the slow-varying
amplitude $\epsilon(z;t)$ of the probe field:
\begin{equation}
\epsilon(z;\Omega)=\int_{-\infty}^{\infty}dt\,%
\mathrm{e}^{i\Omega t}\epsilon(z;t).%
\label{2.2}
\end{equation}
The Fourier component $\epsilon(z;\Omega)$ obeys the following Maxwell equation:
\begin{equation}
\left[-i\frac{\Omega}{c}+\frac{\partial}{\partial z}\right]%
\epsilon(z;\Omega)%
=2\pi i\frac{\bar{\omega}}{c}\,%
\tilde{\chi}(z;\Omega)\,\epsilon(z;\Omega)%
\label{2.3}%
\end{equation}
where $\tilde{\chi}(z;\Omega)$ is the Fourier component of the sample
susceptibility. The spectral dependence of the sample susceptibility is shifted
by the carrier frequency such that:
\begin{equation}
\tilde{\chi}(z;t-t')=\mathrm{e}^{i\bar{\omega}(t-t')}%
\chi(z;t-t')%
\label{2.4}%
\end{equation}
where $\chi(z;t-t')$ is the kernel of the susceptibility operator standardly
defined in time representation and determining response of the polarization
wave at time $t$ on the driving signal wave (\ref{2.1}) considered at retarded
time $t'<t$.

The susceptibility of the medium is then given by the following expansion:
\begin{eqnarray}
\lefteqn{\tilde{\chi}(z;\Omega)=%
-\sum_{n_1=n,n'}\sum_{n_2=n,n'}\frac{1}{\hbar}\left(\mathbf{d e}\right)_{n_1m}^{*}%
\left(\mathbf{d e}\right)_{n_2m}}%
\nonumber\\%
&&\times
\int\frac{d^{3}p}{(2\pi\hbar)^3}n_0(z)%
f_0(\mathbf{p})%
\nonumber\\%
&&\times
G_{n_1n_2}^{(--)}(\mathbf{p}_{\bot},p_z+\hbar\bar{k};\hbar(\bar{\omega}+\Omega)+\frac{p^2}{2 m})%
\label{2.5}%
\end{eqnarray}
We have used the following notations: the transition matrix
elements of the dipole operator $\mathbf{d}$ are projected onto
the polarization vector of the probe mode $\mathbf{e}$, $n_0(z)$
is the density distribution of atoms, and $f_0(\mathbf{p})$ is
their momentum distribution, normalized such as
\begin{equation}
\int\frac{d^{3}p}{(2\pi\hbar)^3}f_0(\mathbf{p})=1.
\label{2.6}%
\end{equation}
The spectral behavior of the susceptibility (\ref{2.5}) is
determined by the spectral properties of the contributing Green's
functions $G_{n_1n_2}^{(--)}$, describing the dynamics of the
upper atomic states dressed by the control mode. The derivation
details and precise definitions of the Green's function, matrix
elements etc. are given in Appendix \ref{A}. We point out here
that these functions for all possible $n_1=n,n'$ and $n_2=n,n'$
built the $2\times 2$ matrix so that the expansion (\ref{2.5})
consists of four terms. Two diagonal terms asymptotically (for
highly resolved hyperfine structure) reproduce the
$\Lambda$-scheme results for either upper state $|n\rangle$ or
$|n'\rangle$. The two others terms, which are interference terms,
significantly modify the predictions of the $\Lambda$-scheme
approximation, as we will show in the following.

\section{Spectral behavior of the sample susceptibility}\label{SectionIII}

In the case of monochromatic excitation, it is convenient to
analyze the Fourier component of the dielectric susceptibility as
a function of either its carrier frequency $\bar{\omega}$ or its
detuning $\bar{\Delta}=\bar{\omega}-\omega_0$ from the atomic
resonance frequency $\omega_0$. In this section we present our
calculation of the sample susceptibility
$\chi=\chi(\bar{\Delta})=\chi'(\bar{\Delta})+i\chi''(\bar{\Delta})$,
which were obtained for an homogeneous medium consisting of
${}^{133}$Cs atoms. We assume a typical configuration of magneto
optical trap (MOT) such that the effects of atomic motion on the
time scale of a few microseconds can be disregarded.

In the following, the susceptibility given by Eq.(\ref{2.5}) will
be scaled in units of $n_0\lambdabar^3$, where $n_0$ is the
density of atoms and $\lambdabar=\lambda /2\pi$ . In typical MOT
conditions, the atomic gas is dilute and $n_0\lambdabar^3\ll 1$
but the sample can be optically thick if $n_0\lambdabar^2L\gg 1$,
where $L$ is the sample length. It is very important to underline
that the transition matrix elements $V_{nm'}$ and $V_{n'm'}$,
contributing to the Green's functions given by Eqs. (\ref{a.4}),
(\ref{a.5}), cannot be considered as independent parameters. They
are expressed by one reduced matrix element multiplied by two
different but dependent algebraic factors, which are determined by
an electronic and nuclear coupling scheme in the upper hyperfine
states and by the polarization of the exciting field. As an
external characteristic of the coupling strength with the control
mode, we use the Rabi frequency $\Omega_c=2|V_{nm'}|/\hbar$
defined with respect to the lower energy transition. The results
of our calculations will be compared with the usual
$\Lambda$-scheme where only state $|n\rangle$ is taken into
consideration.

\begin{figure}
\includegraphics[width=0.54\textwidth]{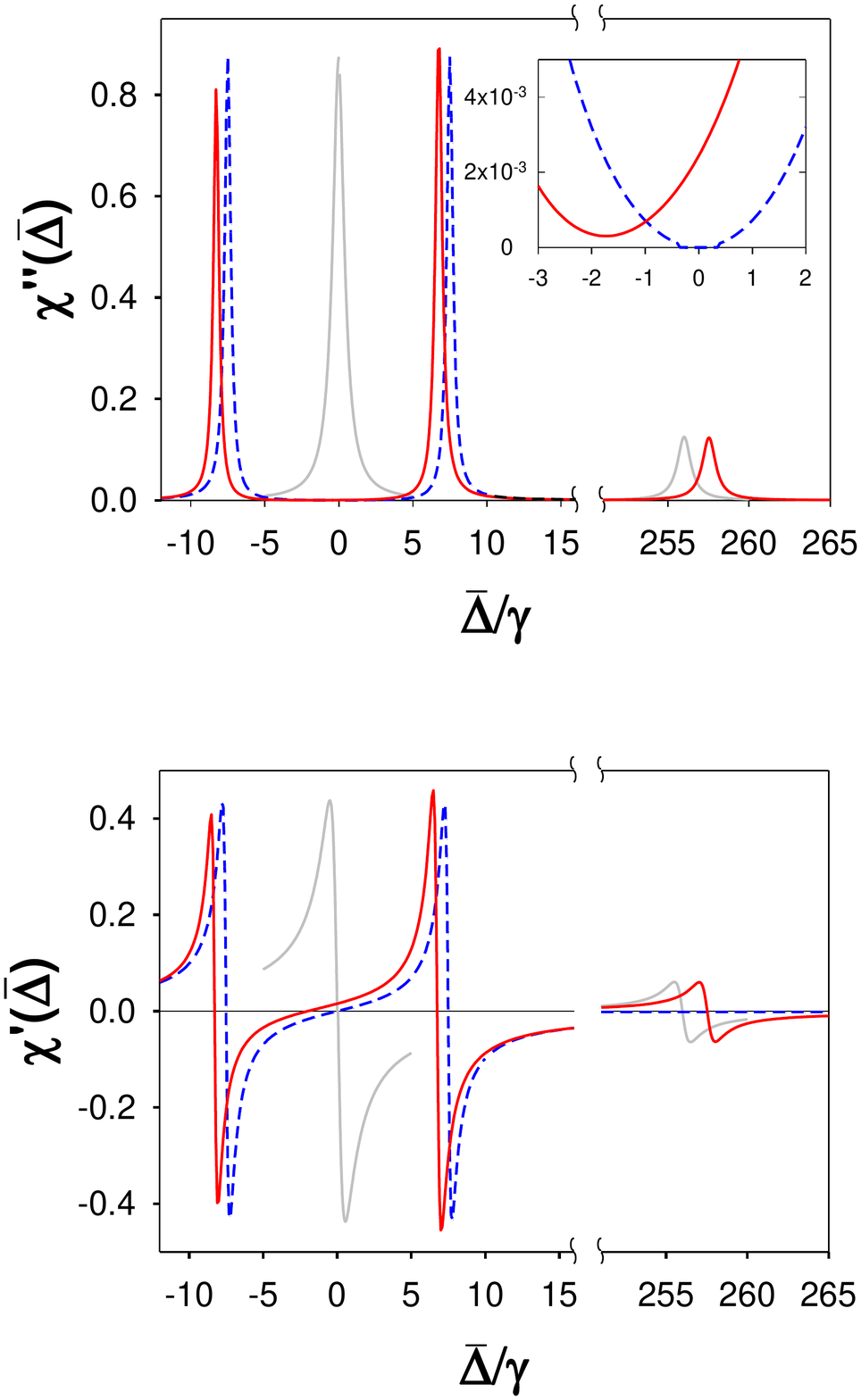}%
\caption{(Color online) Absorption part (upper panel) and
dispersion part (lower panel) of the sample susceptibility in the
presence of the control field tuned in resonance with the atomic
transition ($\Delta=0$), for a Rabi frequency $\Omega_c=15\gamma$.
The susceptibility components are scaled by $n_0\lambdabar^3$. Red
solid curves show the result of the exact calculations and blue
dashed curves are plotted for the usual $\Lambda$-scheme
approximation. For reference, grey curves give the profiles of the
atomic resonances without control field. As shown in the inset of
the upper panel, the EIT resonance is shifted to the red and the
medium looses its perfect transparency at this point.}
\label{fig3}%
\end{figure}

Figure \ref{fig3} gives the spectral dependencies for imaginary
(absorption) and real (dispersion) components of the
susceptibility $\chi(\bar{\Delta})$ (scaled in units of
$n_0\lambdabar^3$) for a Rabi frequency $\Omega_c=15\gamma$ and
for the control field on resonance with the atomic transition
$\Delta=0$. In these conditions, there is no light absorption in
the medium. The imaginary part of the susceptibility is
responsible for the losses caused by incoherent scattering. All
three AT components are shown as well as the doublet approximation
calculated in the three-level $\Lambda$ model when the upper state
$|n'\rangle$ is disregarded. At first sight, the discrepancy
between the exact result and the $\Lambda$-scheme approximation
seems small. However, two important qualitative differences can be
pointed out.

First, in the multilevel configuration, the spectral point, where
due to electromagnetically-induced transparency (EIT) the
absorption falls down to its minimum level, is shifted to the red
and the EIT resonance does not occur for $\bar{\Delta}=\Delta$, as
it always does in the case of the $\Lambda$ model. As clearly
shown in the inset of figure \ref{fig3}, the light shift is
significant and it depends on the Rabi frequency of the control
field. This light shift was first observed by L. Hau and coworkers
in the pioneering experiment on slow light \cite{LHau}. Second,
the medium does not have a perfect transparency at the optimal EIT
point as it also always does for the $\Lambda$-scheme
approximation. In our example this effect seems rather weak but it
is non-vanishing even in the limit of $\Omega_c\to 0$, and it can
become important if the medium has a very large optical depth,
where the losses caused by incoherent scattering can accumulate
\cite{OurD2-paper}.

\begin{figure}
\includegraphics[width=0.52\textwidth]{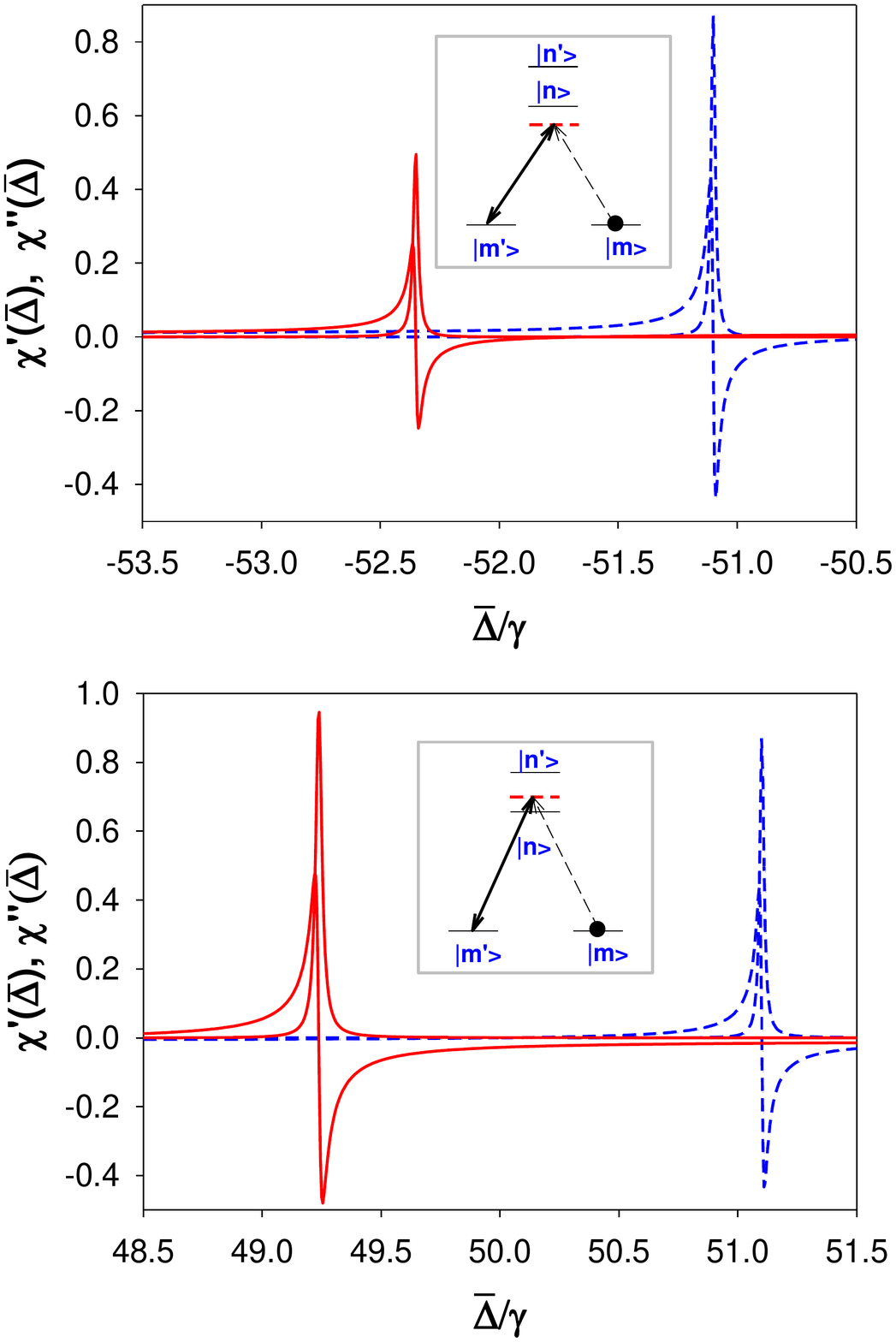}%
\caption{(Color online) Same as in figure \ref{fig3} but for two different detunings. The upper
panel corresponds to $\Delta=-50\gamma$ and the lower panel to $\Delta=50\gamma$ .
Only one component of the AT triplet, which is located near $\bar{\Delta}\sim \Delta$, is shown.}
\label{fig4}%
\end{figure}

In figure \ref{fig4}, the absorption and dispersion parts of the
sample susceptibility are shown for two detunings of the control
mode: $\Delta=-50\gamma$ and $\Delta=50\gamma$. For both
detunings, the lines close to the atomic resonances are only
slightly disturbed by the control field and we show only the part
of the spectral domain where $\bar{\Delta}\sim \Delta$. We compare
the profile of the AT resonance located near the frequency of the
control mode with predictions of the $\Lambda$-scheme
approximation. The dependencies of figure \ref{fig4} show much
larger deviation with the calculations based on the
$\Lambda$-scheme than in the situation of figure \ref{fig3}. This
is a direct consequence of the fact that the detuning $\Delta$ is
comparable with the hyperfine interaction between electronic
angular momentum and nuclear spin.

If the hyperfine interaction was negligible, which asymptotically
occurs when one moves the AT resonance either much lower than
state $|n\rangle$ or much higher than state $|n'\rangle$, then the
considered scheme would be equivalent to the excitation of an atom
with the nuclear spin decoupled from the electronic angular
momentum, which has one-half value. The considered $\Lambda$-type
interaction between the ground state Zeeman sublevels is
impossible in this case. This explains why, when the detuning
$\Delta$ is not negligible with respect to the hyperfine splitting
$\Omega_{HF}$ (here $\Delta=-50\gamma=-\Omega_{HF}/5.12$), the
amplitude of the AT resonance, shown in the upper panel of figure
\ref{fig4} is reduced as compared to the value given by
calculations based on the $\Lambda$-scheme model. On the other
hand, both models show approximately the same narrowing of the
resonance line-width with increasing of $\Delta$. More generally
modifications of the AT resonance in the vicinity of either $D_1$
or $D_2$ transition due to the interplay between the hyperfine
interaction and the optical detunings have a strong impact on
observation of the EIT effect in the hot atomic gas and will be
discussed in details in \cite{OurD2-paper}.

The situation changes if the control mode is tuned between the
hyperfine components as shown in the lower panel of figure
\ref{fig4} for $\Delta=+50\gamma$. One can observe an enhancement
of the AT effect in comparison with predictions of the
$\Lambda$-scheme approximation. This enhancement is even slightly
larger for the dispersion component, as it is more sensitive to
the interference terms contributing into the sample susceptibility
(see Eqs.(\ref{2.5}) and (\ref{a.4})).  As a consequence the
signal pulse passing through the atomic ensemble can undergo
longer delay at the output than it would be expected from the
$\Lambda$-scheme model. This constitutes an interesting advantage
when the Raman process is applied to a quantum memory protocol in
this configuration. This is detailed in the next section.

\section{Application to quantum memory}\label{SectionIV}

As shown before, the Raman  configuration with positive detuning
may lead to an improvement for the storage of a signal pulse. We
evaluate here the performances of a memory in this particular
case, where the control mode is tuned between the upper state
hyperfine levels. In all the following, the Rabi frequency of the
control mode is $\Omega_c=15\gamma$ and it is detuned by
$\Delta=50\gamma$, as in the lower panel of figure \ref{fig4}.

\subsection{Delay of the pulse transport in the atomic sample}

\begin{figure}
\includegraphics[width=0.48\textwidth]{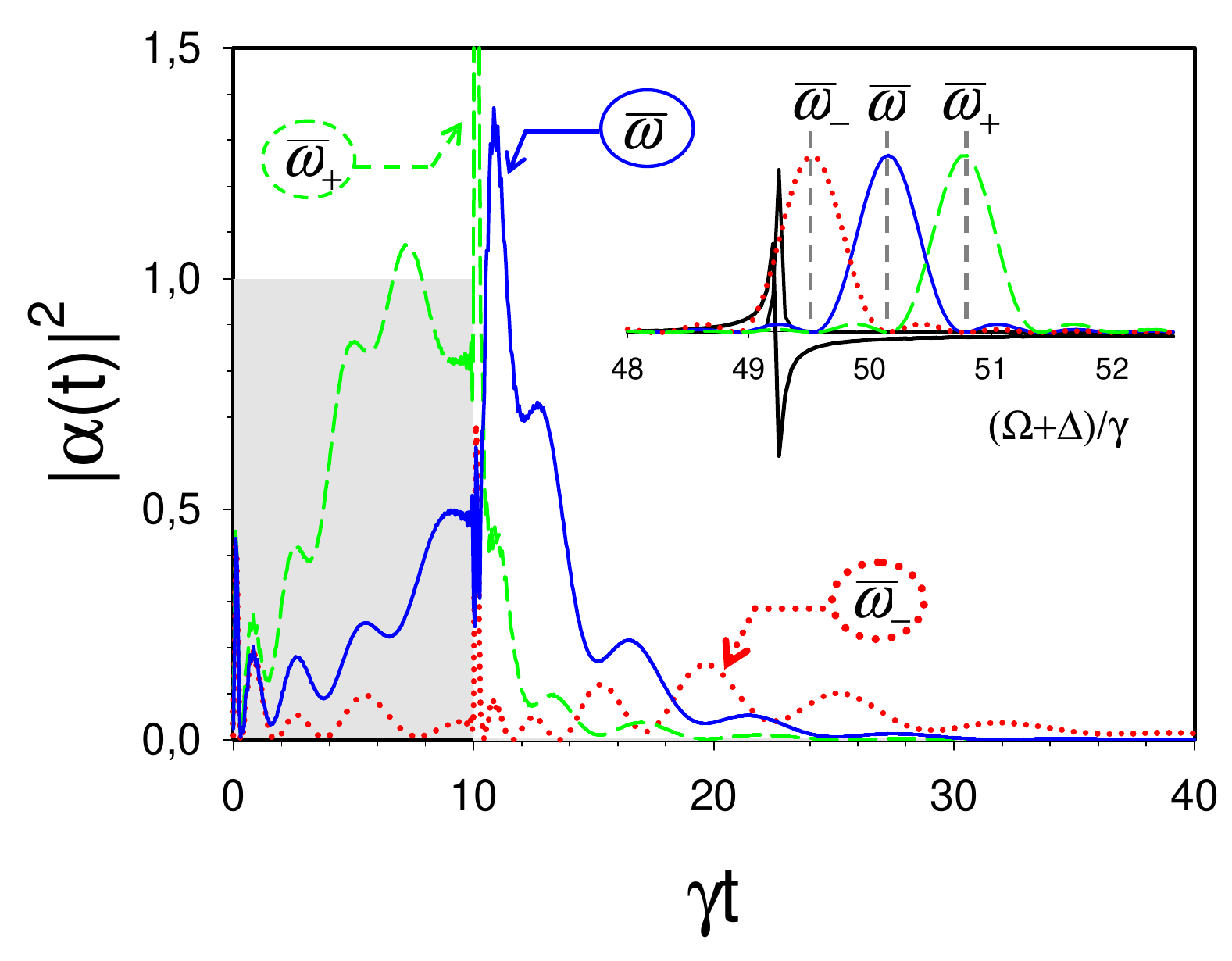}%
\caption{(Color online) Time dependence of the outgoing signal
pulses. The time $t=0$ corresponds to the arrival of the incoming
pulse front to the atomic medium. For the reference, the gray box
gives the outgoing pulse in the absence of atomic medium. Three
pulses the with same duration $T=10\gamma^{-1}$ but different
carrier frequencies $\bar{\omega}_{-}$, $\bar{\omega}$ and
$\bar{\omega}_{+}$ where $\bar{\omega}_{\pm}$ are shifted by $\pm
2\pi/T$, propagate through the optically thick sample with an
optical depth $n_0\lambdabar^2L=50$. The inset shows the pulse
spectrums with respect to the AT resonance. The $\bar{\omega}_{-}$
pulse (red dotted line) experiences the longest delay but also the
largest level of losses due to incoherent scattering. The
$\bar{\omega}_{+}$ pulse (green dashed line) has the lowest level
of losses but also the smallest delay. The central pulse has a
intermediate carrier frequency and corresponds to a balance
between the losses and delay.}
\label{fig5}%
\end{figure}

We consider a coherent pulse with rectangular profile impinging on
the sample. Using Eqs.(\ref{2.5}) and (\ref{2.3}) we can calculate
the shape of the outgoing pulse for various situations. We have
considered here three signal pulses having identical shape but
different carrier frequencies. This approach allows to model a
multimode situation where we make a discrete Fourier expansion
with frequencies that are multiples of $2\pi/T$, where $T$ is the
pulse duration. In the following, we use a dimensionless amplitude
for the signal field denoted by $\alpha(z,t)$. We assume input
pulses with rectangular profile, such that
$\alpha_{\mathrm{in}}(t)=\theta(t)-\theta(t-T)$, where $\theta(t)$
is the step-function, $\theta(t)=1 (t>0)$ or $=0(t<0)$. Figure
\ref{fig5} shows how this pulse is delayed and how its shape is
modified after passing through the atomic medium.

 As shown in the inset of figure \ref{fig5}, the central pulse has a carrier
frequency $\bar{\omega}$ chosen to ensure some balance between
transparency and delay. The two other pulses are shifted by $\pm
2\pi/T$, with $\bar{\omega}_{\pm}=\bar{\omega}\pm 2\pi/T$, which
makes them spectrally orthogonal to the central mode. By
extrapolating these series to infinite number of modes
($\bar{\omega}_q=\bar{\omega}+2\pi q/T$, where $q=0,\pm 1,\pm
2,\ldots$) we can extract the complete set of Fourier modes and
use them for the expansion of an input signal pulse of arbitrary
temporal profile with sharp bound. Such a Fourier description of
an arbitrary pulse of a finite duration $T$ allows to follow the
transformation of each spectral component in the output and could
be important for the multimode quantum information processing.

The delay of the pulse propagating through the optically dense
medium of length $L$ strongly depends on the optical depth. For a
monochromatic signal mode at frequency $\bar{\omega}$ the optical
depth $b(\bar{\omega})$ is given by
\begin{equation}
b(\bar{\omega})=4\pi\chi''(\bar{\omega})\frac{L}{\lambdabar}.
\label{3.1}
\end{equation}
The depth varies from $b(\bar{\omega})\gg 1$ near the points of
the AT resonances to $b(\bar{\omega})\ll 1$ in the transparency
domains. In the $\Lambda$-scheme approximation, the resonance
optical depth is defined in a unique way. In the multilevel
situation the optical depth at resonance depends on the chosen
transition. However it is always closes to $b_0\sim
n_0\lambdabar^2L$, which was taken equal to $50$ for the round of
calculations presented in figure \ref{fig5}.

It can be seen in Figure \ref{fig5} that the shape of the outgoing
pulses is strongly modified as compared to the initial pulse.
First, all three pulses corresponding to $\bar{\omega}$,
$\bar{\omega}_{+}$ and $\bar{\omega}_{-}$ are delayed. The pulse
with frequency $\bar{\omega}_{-}$, the spectrum of which overlaps
the most with the AT peak, spreads over longer times and has a
much longer average delay than the other ones. However, this pulse
has the highest level of losses due to the incoherent scattering.
The pulse with carrier frequency $\bar{\omega}_{+}$ has a lower
level of losses and better preserves the original rectangular
shape, but the delay is quite small.

We can interpret the outgoing signal pulses in Figure \ref{fig5}
in the following way : the part of the pulse going out of the
medium at times smaller than the time duration of the original
pulse, $T=10\gamma^{-1}$, corresponds to direct transmission,
while the tail of the outgoing pulse, which is emitted after the
end of the incoming pulse, corresponds to a potentially stored and
recovered signal. This is a good approximation if the transient
processes associated with switching off/on the control mode as
well as the retardation effects are negligible. The fact that in
off-resonant stimulated Raman scattering the transient processes
only weakly interfere with the transport of the signal pulse was
verified in \cite{MLSSK}.

\subsection{Efficiency of pulse storage and retrieval}

\begin{figure}[htpb!]
\includegraphics[width=0.48\textwidth]{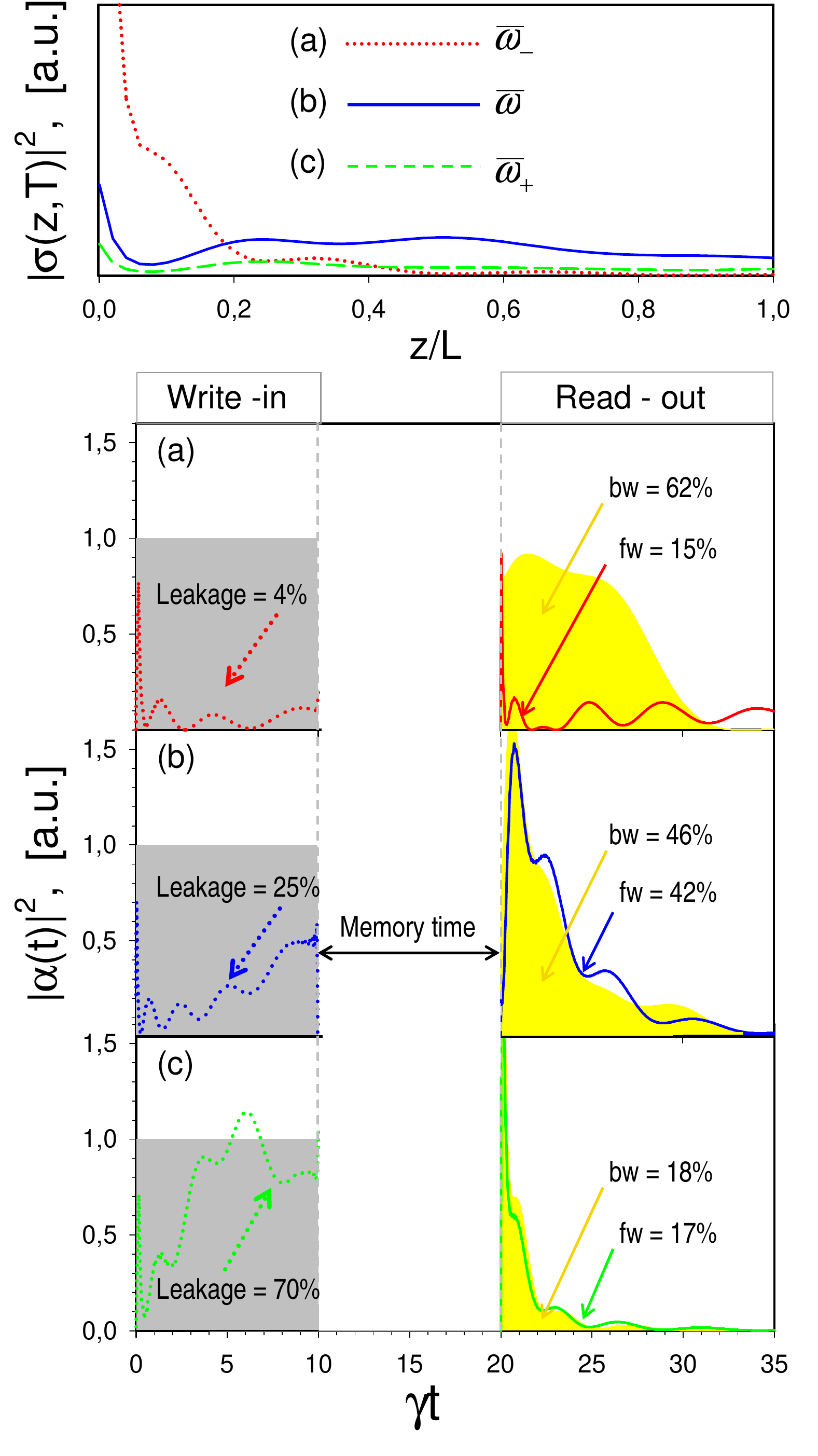}%
\caption{(Color online) Quantum memory. This figure shows the
dynamics of the system for the same parameters as described in
figure \ref{fig5}. Upper panel shows the spin distribution
$\sigma(z,T)$ in the sample at time $T$ ("stored light"). In the
lower panel the signal pulse is retrieved on demand after a
certain memory time via switching on the control field in either
forward (solid lines) or backward (yellow filled area) direction.
The leakage of each pulse, which is the light transmitted during
the write-in stage, is indicated by the dotted line ((a): the
input pulse carrier frequency is $\bar{\omega}_{-}$, (b): the
input pulse carrier frequency is $\bar{\omega}$ ,(c): the input
pulse carrier frequency is $\bar{\omega}_{+}$). The read-out stage
is characterized by the memory efficiency for the forward (fw) or
backward (bw) retrieval.}
\label{fig6}%
\end{figure}

We now turn to the full memory protocol, including storage and
retrieval stages. The control pulse is switched off when the end
front of the incoming signal pulse has entered the sample. After
short transient dynamics, the signal wave packet, which is
localized in the sample, is mostly converted into a standing spin
wave distributed into the whole sample. This spin-wave is given by
the off-diagonal matrix elements (spin coherence) between Zeeman
sublevels $m$ and $m'$ existing in the system at time $T$, which
we denote as $\sigma(z,T)$. To calculate the spin distribution we
have applied the theoretical approach developed in \cite{MKMP} and
modified it for the present case by including the dissipation
processes. The spin distribution is shown in the upper panel of
figure \ref{fig6}.

After a controllable delay, the spin wave can be converted back
into the signal wave packet by sending the second control pulse.
This can be done in either forward or backward directions. In the
first case, the retrieved pulse would reproduce the tail of the
outgoing signal pulse, the propagation of which was interrupted in
the writing stage of the memory protocol. This can be
 seen by comparing the graphs of figures \ref{fig6} and
\ref{fig5}.

The pulse retrieval in the backward direction leads to some
advantages. As discussed in \cite{Gorshkov}, while applying the
time-reversal arguments one can expect a higher efficiency for
backward retrieval after a round of optimization. Also, the shape of
the pulse recovered in the backward direction better reproduces its
original profile. Our numerical simulations confirm this statement.
The spin distribution for the mode $\bar{\omega}_{-}$ indicates that
the signal is mostly stored near the left bound of the sample. Then
it seems more natural to redirect the retrieved pulse in the
backward direction where it would have less absorption and,
according to time-reversal arguments, would be regenerated by the
medium to its original profile. That can be seen in figure
\ref{fig6}, which shows approximately rectangular retrieved profile
for the $\bar{\omega}_{-}$ mode with efficiency more than 60\%.

To conclude this section let us underscore the fact that the
predicted quality of the considered memory channel is quite good.
The obtained results show that in conditions currently attainable
for a MOT the efficiency of the memory protocol could be expected
up to $60\%$. We can expect further improvement in the efficiency
by using the optimization of the temporal profile of the input
pulse, as described in ref. \cite{Gorshkov,Novikova}.

\section{Conclusion}
In this paper we have considered the propagation of a signal pulse
in conditions of stimulated Raman process through a sample of alkali
atoms.  We have shown that the presence of the hyperfine interaction
in the upper state significantly modifies the Autler-Townes effect
observed in the $D_1$-line of such atoms. In particular, we have
demonstrated that it is more efficient to tune the control mode
between the upper state hyperfine sublevels where the stimulated
Raman process is enhanced and provides more effective delay of the
signal pulse.

We have analyzed the dynamics of a signal pulse with initial
rectangular profile and calculated how such a pulse is delayed and
how its shape is modified after passing through the atomic medium.
This dynamics reveals that the outgoing pulse is quite sensitive to
the detuning of the pulse carrier frequency from the Autler Townes
resonance created by the control field. We have used our results to
evaluate the efficiency of a quantum memory protocol based on the
same stimulated Raman process. This has been done both for the
forward and the backward retrieval schemes. We have shown that the
backward retrieval can be more effective and the shape of recovered
pulse can better reproduce its original profile. In particular, our
results show that in conditions currently attainable for a Cesium
magneto-optical trap the efficiency of the memory protocol could be
expected more than $60\%$.

\acknowledgements%
This work was supported by RFBR (Grants No. 10-02-00103, 08-02-91355) and by
the EC under the ICT/FET project COMPAS. A.S. and L.G. acknowledge the
financial support from the charity Foundation "Dynasty" and O.M. from the
Ile-de-France programme IFRAF.

\appendix
\section{The atomic Green's functions and sample susceptibility}\label{A}

The atomic Green's functions and sample susceptibility are found in the second quantization formalism using a non-stationary  diagram technic \cite{Keldysh,LaLfX,MLSSK}. The
Maxwell equation considered in the main text of the paper can be expressed by the
following diagram equation:
\begin{equation}
\scalebox{0.65}{\includegraphics*{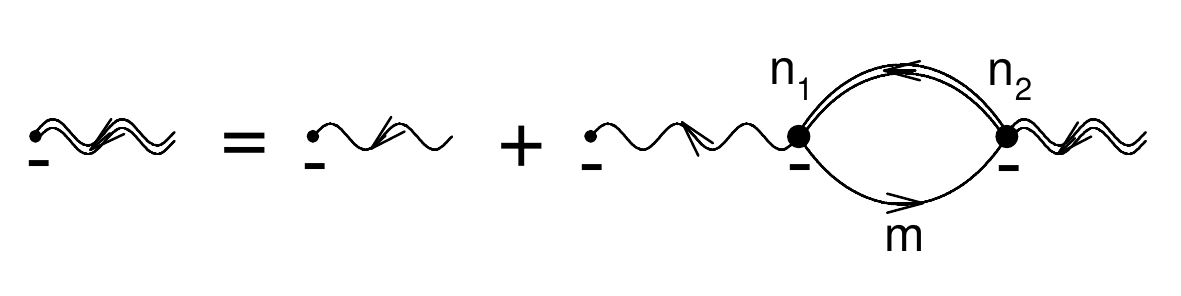}}%
\label{a.1}%
\end{equation}
Here the double wavy single ended lines describe the signal field amplitude
dressed by the interaction process shown in figure \ref{fig1} when this field
propagates through the sample. The single double ended wavy line is the
retarded Green's function for light propagating freely in vacuum. The loop
consisted of the atomic Green's functions and the vertices describe the
polarization operator or susceptibility of the sample in response to the probe
field. For the sake of clarity, we show the internal indices $n_1$ and $n_2$
running the quantum numbers of excited atomic states $n,n'$ (figure
\ref{fig1}).

The retarded-type Green's functions of the excited atomic states dressed by
the interaction with the control and vacuum modes and shown as double line in the
polarization operator of Eq.(\ref{a.1}) are expressed by the following Dyson
equations:
\begin{equation}
\scalebox{0.65}{\includegraphics*{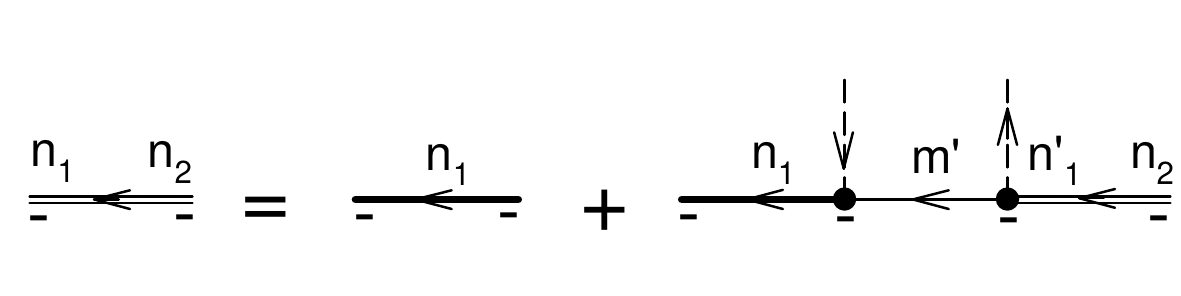}}%
\label{a.2}%
\end{equation}
The outward and inward directed dashed arrows represent the field amplitude of
the control mode and the thick solid line is the retarded-type Green's function
of state $|n_1\rangle$ dressed by interaction with the vacuum modes only. This
function in the first term in the right side should be multiplied by the factor
$\delta_{n_1\,n_2}$ (not shown). It is expressed by the diagram equation
defining the natural decay rate of the excited state:
\begin{equation}
\scalebox{0.65}{\includegraphics*{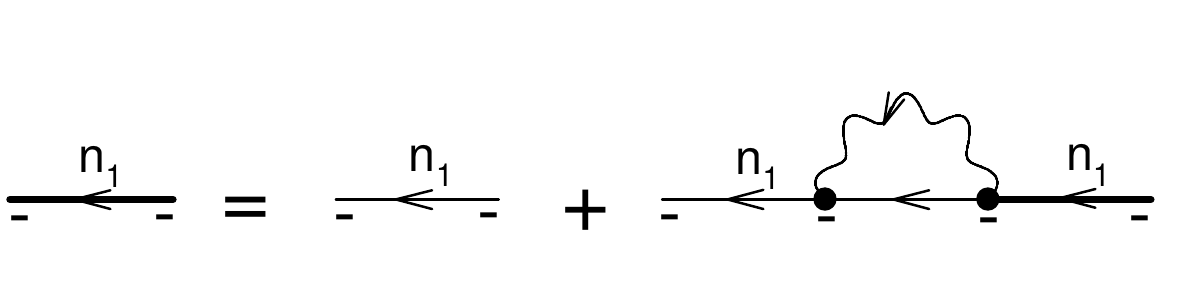}}%
\label{a.3}%
\end{equation}
The thin solid and wavy lines are respectively non-perturbed vacuum Green's
functions of the atom and field.

The minus signs in these diagrams indicate the causal (time-ordered) character
in averaging of the operators' product. In the case of field the expectation
value is taken over its vacuum state. But in the case of atoms the averaging is
taken over the initial state of atomic system. As a consequence, if in the
above diagrams the time naturally rises from right to left, the backward
directed thin line in the polarization operator (\ref{a.1}) is proportional to
the density of atoms in the initially populated state $|m\rangle$. The diagrams
(\ref{a.1})-(\ref{a.3}) can be interpreted as a graphical solution of the Bloch
equations in the first order response to the signal field.

If the control mode is monochromatic then after Fourier transform the integral
equations (\ref{a.2}) can be written in a form of algebraic equations. These
equations give us the set of four Green's functions
$G_{n_1n_2}^{(--)}(\mathbf{p},E)$ defined in the reciprocal space. These
functions contribute into equation (\ref{a.1}) with the "on shell" energy
argument, which is given by $E=\hbar(\bar{\omega}+\Omega)+{p^2}/{2 m_0}$ i. e.
is expressed by the kinetic energy of atom ${p^2}/{2 m_0}$ ($\mathbf{p}$ and
$m_0$ are atomic momentum and mass respectively; the internal ground state
energy is considered as zero level) plus the energy of a photon from the signal
field $\hbar(\bar{\omega}+\Omega)$. Let us point out that because of the
momentum conservation in the polarization operator the momentum of excited atom
$\mathbf{p}'$ is shifted from $\mathbf{p}$ by the momentum of absorbed signal
photon such that $p_z'=p_z+\hbar\bar{k}$.

The Green's functions have the following analytical expression
\begin{eqnarray}
G_{nn}^{(--)}(\mathbf{p},E)&=&\hbar\left\{E-\frac{p^2}{2 m_0}-E_n+i\hbar\frac{\gamma}{2}\right.%
\nonumber\\%
&&\left.-\frac{|V_{nm'}|^2\left[E-\frac{p^2}{2 m}-E_{n'}+i\hbar\frac{\gamma}{2}\right]}%
{\left[E-E_{n'+}(\mathbf{p},\omega)\right]\left[E-E_{n'-}(\mathbf{p},\omega)\right]}\right\}^{-1}%
\nonumber\\%
G_{n'n}^{(--)}(\mathbf{p},E)&=&\frac{V_{n'm'}V_{nm'}^{*}}%
{\left[E-E_{n'+}(\mathbf{p},\omega)\right]\left[E-E_{n'-}(\mathbf{p},\omega)\right]}%
\nonumber\\%
&&\times G_{nn}^{(--)}(\mathbf{p},E)%
\nonumber\\%
&&\ldots%
\label{a.4}%
\end{eqnarray}
Two other functions can  be similarly written via transposing the indices
$n\Leftrightarrow n'$.

The following notation is used: $V_{nm'}=\left(\mathbf{d E}\right)_{nm'}$,
$V_{n'm'}=\left(\mathbf{d E}\right)_{n'm'}$ are the matrix elements of
interaction with the control field; $\mathbf{E}$ is the complex amplitude of
the field's positive frequency component
$\mathbf{E}^{(+)}(\mathbf{r},t)=\mathbf{E}\exp(-i\omega t + i\mathbf{k r})$;
$E_n$, $E_{n'}$ are non-perturbed energies of the excited states; $\gamma$ is
the natural decay rate of the excited state. The quasi-energies in the
denominators of Eqs.(\ref{a.4}) contribute to the excitation spectrum dressed
by interaction with vacuum modes and control field while the hyperfine
interaction is infinitely strong. They are given by
\begin{eqnarray}
\lefteqn{E_{n\pm}(\mathbf{p},\omega)=E_{m'}+\frac{\mathbf{p}^2}{2m_0}%
+\frac{\hbar}{2}\left[\omega-\frac{\mathbf{k
p}}{m_0}+\omega_{nm'}-i\frac{\gamma}{2}\right]}%
\nonumber\\%
&&\pm\left[|V_{nm'}|^2+\frac{\hbar^2}{4}%
\left(\omega_{nm'}-\omega+\frac{\mathbf{k
p}}{m_0}-i\frac{\gamma}{2}\right)^2\right]^{1/2}%
\label{a.5}%
\end{eqnarray}
and similar for $n'$. Here $E_{m'}$ is the energy of state $|m'\rangle$, which
for the system of degenerate Zeeman sublevels shown in figure \ref{fig1}
coincides with the energy of state $|m\rangle$, such that $E_{m'}= E_{m}=0$.
Quasi-energies (\ref{a.5}) are convenient intermediate parameters describing
the AT structure if only one state either $|n\rangle$ or $|n'\rangle$ is taken
into consideration. But actual location of the AT poles is described by full
susceptibility of the sample given by Eq.(\ref{2.5}), where all four Green's
functions equally contribute.

\end{document}